\documentclass[conference]{IEEEtran}
\IEEEoverridecommandlockouts

\usepackage{amsmath,amssymb,amsfonts}
\usepackage{algorithmic}
\usepackage{graphicx}
\usepackage[nolist]{acronym}
\usepackage{textcomp}
\usepackage{xcolor}
\usepackage{siunitx}
\usepackage{float}
\usepackage{multirow}
\usepackage{adjustbox}
\usepackage{makecell}
\usepackage{tabularx}
\usepackage{cellspace}
\usepackage{ragged2e} 
\usepackage{hyperref}

\usepackage[style=ieee, backend=biber, natbib=true,  maxbibnames=1, minbibnames=1]{biblatex}

\newboolean{blindreview}
\setboolean{blindreview}{false}

\usepackage{fancyhdr}

\fancypagestyle{firstpage}{%
	\fancyhf{}%
	\fancyhead[C]{\normalsize To appear at 2025 IEEE International Symposium on Circuits and Systems, May 25-28 2025, London, United Kingdom}
	
	\fancyfoot[C]{\footnotesize \copyright2025 IEEE. Personal use of this material is permitted. Permission from IEEE must be obtained for all other uses, in any current or future media, including reprinting/republishing this material for advertising or promotional purposes, creating new collective works, for resale or redistribution to servers or lists, or reuse of any copyrighted component of this work in other works.}
}

\begin{document}

\begin{acronym}[placeholder]		
	\acro{SoCs}{System-on-Chips}
	\acro{SoC}{System-on-Chip}
	\acro{SV}{SystemVerilog}
	\acro{SV-UVM}{SystemVerilog-Universal Verification Methodology}
	\acro{ECCs}{Error Correction Code}
	\acro{ECC}{Error Correction Code}
	\acro{SECDED}{Single-Error Correction, Double-Error Detection}
	\acro{IPs}{Intellectual Properties}
	\acro{HDLs}{Hardware Description Languages}
	\acro{HVLs}{Hardware Verification Language}
	\acro{CRV}{Constrained Random Verification}
	\acro{CSV}{Comma Separated Values}
	\acro{BFMs}{Bus Functional Models}
	\acro{DUVs}{Designs Under Verification}
	\acro{DUV}{Design Under Verification}
	\acro{SoC}{System-on-Chip}
	\acro{FV}{Formal Verification}
	\acro{SAR}{Successive Approximation Register}
	\acro{acronym}{full name}
	\acro{GPI}{General Purpose Interface}
	\acro{MDV}{Metric Driven Verification}
	\acro{ML}{Machine Learning}
	\acro{HDL}{Hardware Description Language}
	\acro{HVL}{Hardware Verification Language}
	\acro{ADC}[ADC]{Analog-to-Digital Converter}
	\acro{I2C}{Inter-Integrated Circuit}
	\acro{ALU}{Arithmetic Logic Unit}
	\acro{ASIC}{Application-Specific Integrated Circuit}
	\acro{AI}{Artificial Intelligence}
	\acro{IP}{Intellectual Property}
	\acro{Cocotb}{Coroutine-based cosimulation testbench}
	\acro{TLM}{Transaction Level Modeling}
	\acro{UVM}{Universal Verification Methodology}
	\acro{DUT}{Design Under Test}
	\acro{BFM}{Bus Functional Model}
	\acro{EDA}{Electronic Design Automation}
	\acro{RAL}{Register Abstraction Layer}
	\acro{VPI}{Verilog Procedural Interface}
	\acro{VHPI}{VHDL Procedural Interface}
	\acro{FPGA}{Field Programmable Gate Array}
	\acro{PyUVM}{Python Universal Verification Methodology}
\end{acronym}

\title{Optimizing Coverage-Driven Verification Using Machine Learning and PyUVM: A Novel Approach\\
	\thanks{This work has been developed in the project VE-VIDES (project label 16ME0243K) which is partly funded within the Research Programme ICT 2020 by the German Federal Ministry of Education and Research (BMBF)}
}

\ifthenelse{\boolean{blindreview}}{}{
	\author{\IEEEauthorblockN{Suruchi Kumari}
		\IEEEauthorblockA{Infineon Technologies \\
			Dresden, Germany \\
			Suruchi.Kumari@infineon.com}
		\and
		\IEEEauthorblockN{Deepak Narayan Gadde}
		\IEEEauthorblockA{Infineon Technologies \\
			Dresden, Germany \\
			Deepak.Gadde@infineon.com}
		\and
		\IEEEauthorblockN{Aman Kumar}
		\IEEEauthorblockA{Infineon Technologies \\
			Bangalore, India \\
			Aman.Kumar@infineon.com}
	}
}

\maketitle
\thispagestyle{firstpage}

\begin{abstract}
The escalating complexity of \ac{SoC} designs has created a bottleneck in verification, with traditional techniques struggling to achieve complete coverage. Existing techniques, such as \ac{CRV} and coverage-driven methodologies, rely on time-consuming and redundant simulation regression, leading to higher verification costs and longer time-to-market due to the manual effort required to adjust constraints and drive the stimuli to achieve coverage objectives. To address this challenge, we propose a novel methodology that leverages supervised \ac{ML} to optimize simulation regressions, resulting in reduced simulation run-time and the number of test simulations required to achieve target coverage goals. We also investigate and compare the effectiveness of various supervised learning algorithms from scikit-learn \cite{scikit-learn}. Our results demonstrate that these algorithms can achieve at least 99\% coverage regain (\ref{eq:coverage_regain}) with significantly reduced simulation cycles. We utilize \ac{PyUVM} over \ac{SV-UVM} for testbench creation, enabling simpler constructs using Python and facilitating the reuse of existing \ac{ML} libraries. Our methodology is applied to three diverse designs, and our results show that it can significantly reduce verification costs, manual efforts, and time-to-market, while enhancing verification productivity and completeness, by automating the testbench update process and achieving target coverage goals.
\end{abstract}


\section{Introduction}\label{sec:intro}
The rapid advancement of semiconductor technology has enabled the integration of multiple functionalities and features into a single \ac{SoC}. However, design verification remains a significant bottleneck in \ac{SoC} development, consuming a substantial portion of the overall project time and resources. According to a recent study by the Wilson Research Group, verification accounts for around \SI{60}{\percent} of the overall project time, highlighting the need for more efficient verification methodologies \cite{VerStudy_22}.

The simulation-based design verification is a well-established and powerful technique that leverages \ac{CRV} and coverage-driven methodologies, which typically use simulation regressions in the end. The goal is to improve verification objectives, including coverage, by repeating test simulations with random seeds. However, these classical simulation regressions often involve redundant test simulations, which does not guarantee an improvement in the total coverage. This process utilizes significant simulation resources and time to manually adjust constraints, leading to an increase in verification costs and time-to-market for \ac{SoC} products. Additionally, the generation of huge amounts of data in simulation regression makes design verification an ideal field for \ac{ML} application.

Numerous initiatives have been made to optimize functional verification through the use of \ac{ML} techniques, as reviewed in the surveys \cite{article} and \cite{siemens_whitepaper}. Most of these studies focused on functional coverage improvement, simulation speedup, and reducing the test simulations in the simulation regressions. This paper presents a novel approach to addressing the high number of test simulations and longer run-time in classical simulation regression, while meeting the target coverage goals. This study aims to promote the adoption of automation methodologies that utilize data science and ML, showcasing the performance and comparing various supervised models.

The contributions of this work are as follows:
\begin{itemize}
	\item Novel simulation regression optimization using supervised learning (Section \ref{sec:implementation})
	\item \ac{PyUVM}-based testbench creation for \ac{DUVs} (Section \ref{tbcreation})
	\item \ac{ML} environment integration into simulation environment (Section \ref{datacnp}, \ref{mlp})
	\item Automatic testbench updates with ML-predicted constraints, sequences, and tests (Section \ref{automatictb})
\end{itemize}

\section{Essential Background}\label{sec:background}
This section gives an overview of the relevant techniques used in our work.
\subsection{Python-based verification}
\begin{figure}
	\centering
	\includegraphics[width=0.5\linewidth]{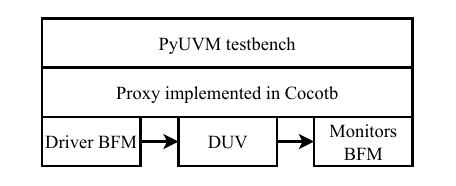} 
	\caption{Proxy-driven PyUVM testbench}
	\label{pyuvmbfm}
\end{figure}

Python is a widely-used, high-level programming language employed across various industries. Its extensive standard library and ecosystem support a multitude of existing libraries. To support the functionalities from SystemVerilog and \ac{UVM}, \ac{Cocotb} \cite{cocotb} is introduced as an open-source Python-based verification environment. It provides synchronous logic, connects the \ac{DUV} and Python testbench using \ac{GPI}. \\Building upon \ac{Cocotb}, \ac{PyUVM} is a Python library that implements \ac{UVM} 1.2 IEEE specification in Python using its high-level language features \cite{salmei_themperek}. As shown in Fig. \ref{pyuvmbfm}, the testbench software is the \ac{PyUVM} testbench, the proxy is implemented in \ac{Cocotb}, where \ac{Cocotb} connects Python to simulators through \ac{VPI} and \ac{VHPI}. Hence, both Python and the \ac{DUV}, which runs in the simulator, share the same proxy interface. The transactions from the testbench are called using Python coroutines such as driver and monitors \ac{BFMs}, which also ensures the \ac{DUV} is not busy.
PyVSC \cite{Ballance} is a Python library that enables constrained randomization and functional coverage,
similar to SystemVerilog. Python does not have in-built support for coverage constructs.
Hence, PyVSC has been used in conjunction with PyUVM to enable these features in this research
work.
\subsection{Machine Learning}
Supervised \ac{ML} is used to obtain trained models to make predictions, which are further utilized for the \ac{ML}-optimized simulation regression. Table \ref{analogy} shows the analogy used in our work for \ac{ML} modeling.

\begin{table}[htbp]
	\setcellgapes{0.5pt}
	\makegapedcells
	\renewcommand{\arraystretch}{1.3}
	\caption{ML modeling terms and their analogy in our work}
	\label{analogy}
	\footnotesize
	\centering
	\resizebox{\columnwidth}{!}
	{
\begin{tabular}{lll}
	\hline
	\textbf{ML componets}                                            & \textbf{Definition}                                                                                    & \textbf{Analogy in our work}                                                  \\ \hline
	Dataset                                                          & Data used for training/testing ML models                                                               & \begin{tabular}[c]{@{}l@{}}I/P and coverbin status \\ (hit/miss)\end{tabular} \\ \hline
	\begin{tabular}[c]{@{}l@{}}Independent \\ variables\end{tabular} & \begin{tabular}[c]{@{}l@{}}Features extracted from dataset and \\ used for O/P prediction\end{tabular} & O/P coverbin                                                                  \\ \hline
	\begin{tabular}[c]{@{}l@{}}Dependent \\ variables\end{tabular}   & O/P values/labels to be predicted                                                                      & I/P stimuli                                                                   \\ \hline
\end{tabular}%
	}
	\vspace{-0.4cm}
\end{table}

\subsection{Performance metrics}
\label{pm}
To evaluate the proposed methodology, three performance metrics \ref{eq:optimization_runs}, \ref{eq:optimization_runtime}, and \ref{eq:coverage_regain} have been defined, referred from \cite{gadde2024improvingsimulationregressionefficiency}.
\begin{equation}
	\scriptsize\textit{Optimization in test runs (x)} = \frac{\textit{Number of test runs in original regression}}{\textit{Number of test runs in ML-optimized regression}}
	\label{eq:optimization_runs}
\end{equation}
\begin{equation}
	\scriptsize\textit{Optimization in run-time (x)} = \frac{\textit{Simulation run-time of original regression}}{\textit{Simulation run-time of ML-optimized regression}}
	\label{eq:optimization_runtime}
\end{equation}
\begin{equation}
	\scriptsize\textit{Coverage regain (\%)} = \frac{\textit{Coverage of ML-optimized regression}}{\textit{Coverage of original regression}} 
	\label{eq:coverage_regain}
\end{equation}

\section{Related Work}\label{sec:designs}
\begin{table}[htbp]
	\setcellgapes{1pt}
	\makegapedcells
	\renewcommand{\arraystretch}{1.3}
	\caption{Comparison of our work with other relevant works}
	\label{relatedwork}
	\footnotesize
	\centering
	\resizebox{\columnwidth}{!}
	{
\begin{tabular}{cllcll}
	\hline
	\textbf{Work}                                                & \textbf{Year} & \textbf{ML Approach}                                                    & \textbf{\begin{tabular}[c]{@{}c@{}}Verification \\ Method\end{tabular}} & \textbf{Training Data}                                                                    & \textbf{Contribution} \\ \hline
	\cite{sarath}                                   & 2018          & Linear, ANN                                                             & SV-UVM                                                                  & \begin{tabular}[c]{@{}l@{}}Input stimuli, \\ coverbin status\end{tabular}                 & SS, TR                \\ \hline
	\cite{kulkarni}                                 & 2019          & \begin{tabular}[c]{@{}l@{}}Various supervised\\ algorithms\end{tabular} & SV-UVM                                                                  & \begin{tabular}[c]{@{}l@{}}Constraints or knobs \\ in DUV\end{tabular}                    & FCC, TR               \\ \hline
	\cite{gaur}                                     & 2019          & ANN                                                                     & SV-UVM                                                                  & I/P and O/P from DUV                                                                      & SS                    \\ \hline
	\cite{varambally2020optimising}                 & 2020          & ANN                                                                     & \begin{tabular}[c]{@{}c@{}}Python\\ Cocotb\end{tabular}                 & \begin{tabular}[c]{@{}l@{}}I/P and O/P from DUV, \\ coverage and test status\end{tabular} & FCC                   \\ \hline
	\cite{ghany}                                    & 2021          & \begin{tabular}[c]{@{}l@{}}ANN, DNN, \\ DT, SVR\end{tabular}            & SV-UVM                                                                  & I/P and O/P from DUV                                                                      & TR                    \\ \hline
	\cite{blackmore}                                & 2022          & NN                                                                      & SV-UVM                                                                  & Test configurations                                                                       & TR                    \\ \hline
	\cite{plucinski}                                & 2023          & NN                                                                      & SV-UVM                                                                  & \begin{tabular}[c]{@{}l@{}}Test parameters and \\ I/P to DUV\end{tabular}                 & SS                    \\ \hline
	\cite{gadde2024efficientstimuligenerationusing} & 2024          & RL                                                                      & SV                                                                      & I/P and O/P code coverage                                                                 & CCC                   \\ \hline
	Our work                                                     & 2024          & \begin{tabular}[c]{@{}l@{}}Various supervised\\ algorithms\end{tabular} & PyUVM                                                                   & \begin{tabular}[c]{@{}l@{}}I/P and coverbin status \\ (hit or not hit)\end{tabular}       & TR, SS                \\ \hline
\end{tabular}
	}
	\begin{center}
		\vspace{1ex}
		\justifying
		\tiny  Notes:- ANN: Artificial Neural Networks, SVR: Support Vector Regression, KNN: K-Nearest Neighbors, SVM: Support Vector Machine, RL: Reinforcement Learning, DNN: Deep Neural Networks, DT: Decision Tree; RF: Random Forest, NN: Neural Networks, SS: Simulation Speedup, TR: Tests Reduction, FCC: Functional Coverage Convergence, CCC: Code Coverage Convergence
	\end{center}
	\vspace{-0.85cm}
\end{table}

In recent years, there has been a growing interest in using \ac{ML} for verification processes to improve functional coverage. Several studies have explored the application of ML algorithms in autonomously updating constraints, classifying inputs, improving test generation, and selecting novel tests. For instance, Ambalakkat et al. \cite{sarath} proposed a methodology to autonomously update constraints using multiple \ac{ML} algorithms, achieving faster coverage closure. Gaur et al. \cite{gaur} classified inputs as randomizable or not using switching probability of output as a metric, and applied it to the Physical Medium Attachment (PMA) block of High-Speed Serial Interface (HSSI). Kulkarni \cite{kulkarni} improved functional coverage using \ac{ML} methods on the Modiﬁed-Exclusive-Shared-Invalid InterSection Controller (MESI ISC) design. Varambally and Sehgal \cite{varambally2020optimising} proposed a software-based methodology on an open-source platform, achieving significant reduction in simulation iterations. Other studies \cite{ghany}, \cite{blackmore}, and \cite{plucinski} have also demonstrated the effectiveness of ML-based approaches in test selection, prediction of input stimuli, and functional coverage. The authors in \cite{gadde2024efficientstimuligenerationusing} utilized reinforcement learning to converge code coverage. Table \ref{relatedwork} shows the comparison of these studies with our work.

However, all these studies mostly utilized SV-UVM, which obstructs on-the-fly collection of data for \ac{ML} modeling. Additionally, the integration of the \ac{ML} environment with the simulation environment is tedious and manual. Most of the approaches/frameworks are not scalable to complex designs and only contribute to either simulation speedup or coverage improvement. In contrast, our work proposes a design-agnostic method that utilizes Python-based verification, easing the integration to the \ac{ML} environment and allowing on-the-fly data collection. We not only focus on achieving target coverage but also simulation speedup and reduction of test simulations while attaining it.

\section{Proposed Methodology}\label{sec:implementation}
\begin{figure}
	\centering
	\includegraphics[width=0.7\linewidth]{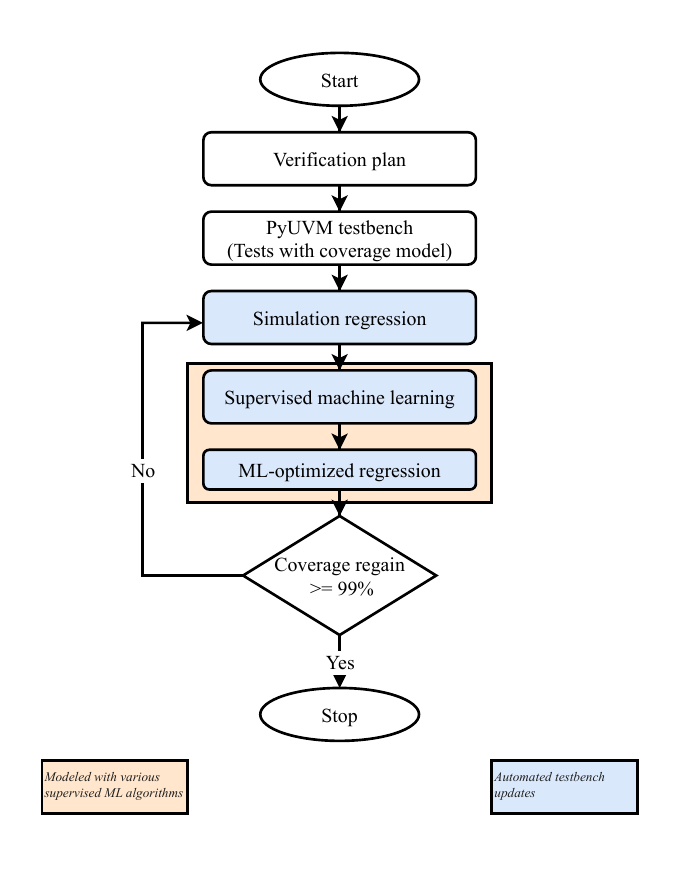} 
	\caption{Proposed methodology}
	\label{proposedmethod}
\end{figure}

The proposed methodology flow, illustrated in Fig. \ref{proposedmethod}, commences with the development of a comprehensive verification plan derived from the design specification. This plan encompasses all verification scenarios and cover items that must be achieved during design verification. Subsequently, a \ac{PyUVM} testbench is created, incorporating functional tests and coverage models defined using PyVSC. The simulation of these tests generates coverage information, which is then merged using PyUCIS to obtain overall coverage \cite{Pyucis}. The application of data science and supervised \ac{ML} techniques to this data enables the creation of models using various \ac{ML} algorithms. These models are utilized to produce an \ac{ML}-optimized regression. A comparative analysis of the functional coverage between the original and \ac{ML}-optimized regressions for various \ac{ML} algorithms is performed. If the coverage regain in the optimized regression reaches \SI{99}{\percent} or higher, the flow is terminated. Otherwise, the flow reverts to the simulation regression step, and the process is repeated until the desired coverage is achieved.
\subsection{Testbench creation and simulation regression}
\label{tbcreation}
\ac{DUVs} utilized in this study are a 32-bit \ac{ALU}, \ac{ADC}, and \ac{SECDED} \ac{ECC}. The cover bins for the \ac{ALU} and \ac{ECC} are deliberately defined to be challenging to hit, while the ADC is an analog-mixed signal model. This allow a more efficient evaluation of the methodology proposed in this research. The testbench is created using PyUVM and PyVSC. The detailed testbench architecture is also discussed in the work \cite{gadde}. The following considerations are taken into account when coding this script for running simulation regression:
\begin{itemize}
	\item Specify the tests defined in the testbench to be run multiple times
	\item Run each test with a different random seed value, generated using the Python random module
\end{itemize}

\subsection{Data collection and preparation}
\label{datacnp}
\ac{PyUVM} testbenches offer a significant advantage in data collection compared to SystemVerilog-UVM testbenches. During simulation regression, \ac{PyUVM} testbenches can collect sampled values of randomized input stimuli and coverbin hit/miss information at every clock cycle, storing them in a \ac{CSV} file.

\begin{figure}
	\centering
	\includegraphics[width=0.85\linewidth]{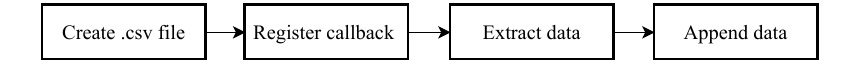} 
	\caption{Data collection}
	\label{datacollection}
\end{figure}

\begin{figure}
	\centering
	\includegraphics[width=0.9\linewidth]{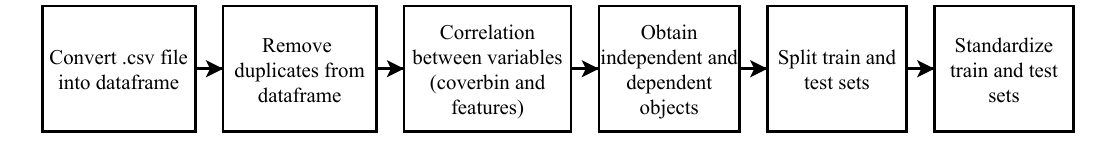} 
	\caption{Data preparation}
	\label{datapreparation}
\end{figure}

PyVSC provides a callback method to collect bin hit/miss information, which is defined in the data collection class and registered in the UVM monitor class. Every clock cycle, the covergroup is sampled, and the information is collected along with the randomized objects. This data is then appended to the CSV file, created at the start of the regression. The process ﬂow of collecting data is presented in Fig. \ref{datacollection}.
As discussed in Fig. \ref{datapreparation}, the collected CSV data is first processed using pandas \cite{pandas} dataframes. Duplicate rows are removed to ensure data quality. Next, the correlation function from the pandas library is applied to determine the dependent variable. The input with the highest correlation value with respect to a particular bin is selected as the dependent variable, while all other inputs are considered independent variables.
\subsection{ML processing}
\label{mlp}
The final dataset is analyzed using supervised machine learning models to identify positive or negative relationships between the variables. These models can also forecast new, unseen observations. This work employs various supervised regression algorithms from the scikit-learn library \cite{scikit-learn}. A separate \ac{ML} model is constructed for each coverbin to train and predict constraints. These models aim to describe the relationships between variables by fitting lines to independent data samples and predicting dependent variable ranges required to hit the coverbins during test simulation.
\subsection{Automatic testbench updation}
\label{automatictb}
To minimize effort, we have implemented an automated solution using Python script to update the testbench. This script take the \ac{ML}-predicted value ranges as input and generate constraints, sequences, and tests for all ML algorithms. This automation enables efficient running of optimized regressions and avoids extra manual effort.

\section{Results}\label{sec:results}
\begin{table}[]
	\setcellgapes{0.5pt}
	\makegapedcells
	\renewcommand{\arraystretch}{1.3}
	\caption{Verification statistics of original and \ac{ML}-optimized regressions}
    \label{vs}
	\footnotesize
	\centering
	\resizebox{\columnwidth}{!}
	{
\begin{tabular}{lcccccccccc}
	\hline
	\multirow{2}{*}{\textbf{\begin{tabular}[c]{@{}l@{}}Verification \\ statistic\end{tabular}}}    & \multirow{2}{*}{\textbf{DUV}} & \multirow{2}{*}{\textbf{Original}} & \multicolumn{8}{c}{\textbf{ML-optimized}}                                                                                       \\ \cline{4-11} 
	&                               &                                    & \textbf{Linear} & \textbf{LASSO} & \textbf{Ridge} & \textbf{SVR} & \textbf{DT} & \textbf{RF} & \textbf{AdaBoost} & \textbf{KNN} \\ \hline
	\multirow{3}{*}{\textbf{\begin{tabular}[c]{@{}l@{}}Number of \\ test runs\end{tabular}}}       & ALU                           & 100                                & \multicolumn{8}{c}{22}                                                                                                          \\ \cline{2-11} 
	& ADC                           & 200                                & \multicolumn{8}{c}{11}                                                                                                          \\ \cline{2-11} 
	& ECC                           & 50                                 & \multicolumn{8}{c}{8}                                                                                                           \\ \hline
	\multirow{3}{*}{\textbf{\begin{tabular}[c]{@{}l@{}}Functional \\ coverage (\%)\end{tabular}}}  & ALU                           & 98.57                              & 99.72           & 99.83          & 99.89          & 99.91        & 99.96       & 99.96       & 99.93             & 99.94        \\ \cline{2-11} 
	& ADC                           & 100.00                             & 99.50           & 94.00          & 96.80          & 96.30        & 95.50       & 98.40       & 98.80             & 98.80        \\ \cline{2-11} 
	& ECC                           & 100.00                             & \multicolumn{8}{c}{100.00}                                                                                                      \\ \hline
	\multirow{3}{*}{\textbf{\begin{tabular}[c]{@{}l@{}}Simulation \\ run-time (sec)\end{tabular}}} & ALU                           & 426.88                             & 175.15          & 171.53         & 174.03         & 180.97       & 183.61      & 185.04      & 175.48            & 185.64       \\ \cline{2-11} 
	& ADC                           & 147.64                             & 3.40            & 3.52           & 3.77           & 3.27         & 2.61        & 2.50        & 2.69              & 2.74         \\ \cline{2-11} 
	& ECC                           & 210.40                             & 7.78            & 8.19           & 9.03           & 8.01         & 8.36        & 8.47        & 8.41              & 8.87         \\ \hline
\end{tabular}
	}
	\vspace{-0.3cm}
\end{table}
\begin{figure}
	\centering
	\includegraphics[width=0.8\linewidth]{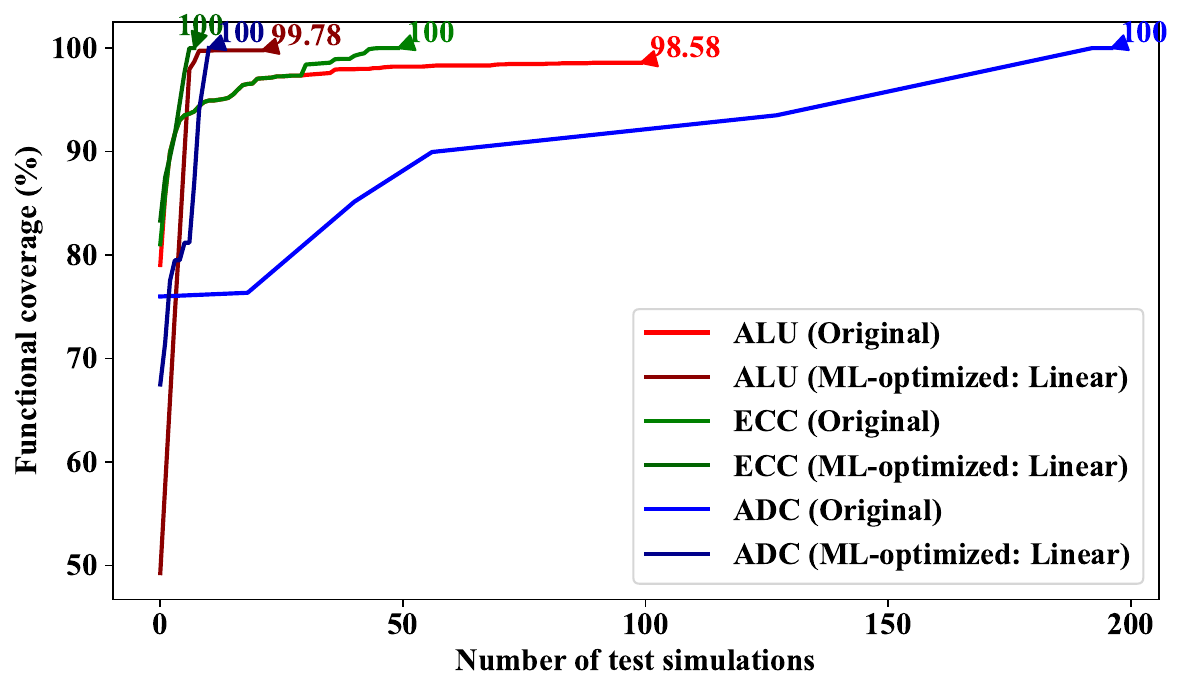} 
	\caption{Functional coverage with respect to test simulations generated by optimized regression}
	\label{funccoverage}
\end{figure}

This section compares verification statistics and performance metrics for original and \ac{ML}-optimized simulation regressions of three \ac{DUVs}, and examines functional coverage advancement with respect to test simulations. The generated \ac{ML}-optimized regressions are random in nature, and when run multiple times with random seeds, they simulate various random scenarios, leading to possible coverage improvement. It is observed that supervised regression algorithms LASSO, Ridge, SVR, DT, RF, AdaBoost, and KNN performed well in optimizing the test simulations and simulation run-times achieving \SI{99}{\percent} coverage regain, except for \ac{ADC} as shown in Table \ref{pm_or}. However, linear regression emerged as the most effective algorithm, attaining \SI{99.5}{\percent} coverage regain for ADC and providing consistent performance for all three \ac{DUVs}.

For each \ac{DUV}, Table \ref{vs} shows that the \ac{ML}-optimized regression significantly reduces the number of test runs, which in turn also reduces the simulation cycles and simulation run-time while achieving higher coverage compared to the original regression. For example, for the ALU, the \ac{ML}-optimized regression requires only 20 test runs and 175 seconds of simulation run-time to achieve \SI{99}{\percent} coverage, compared to 500 test runs and 426 seconds of simulation run-time for the original regression. Table \ref{pm_or} shows that, for example, the optimized regression reduced test simulations by a factor of 18 for \ac{ECC}, optimized run-time by 27 times, and achieved \SI{100}{\percent} coverage regain for linear \ac{ML}-optimized regression. Fig. \ref{funccoverage} illustrates the rapid attainment of maximum functional coverage with fewer test simulations for output-optimized regression using linear models for each \ac{DUV}. 
\begin{table}[]
	\setcellgapes{0.5pt}
	\makegapedcells
	\renewcommand{\arraystretch}{1.3}
	\caption{Performance metrics of ML-optimized regression}
	\label{pm_or}
	\footnotesize
	\centering
	\resizebox{\columnwidth}{!}
	{
\begin{tabular}{lccccccccc}
	\hline
	\multirow{2}{*}{\textbf{\begin{tabular}[c]{@{}l@{}}Performance\\ Metric\end{tabular}}}                    & \multirow{2}{*}{\textbf{DUV}} & \multicolumn{8}{c}{\textbf{Supervised models}}                                                                                  \\ \cline{3-10} 
	&                               & \textbf{Linear} & \textbf{LASSO} & \textbf{Ridge} & \textbf{SVR} & \textbf{DT} & \textbf{RF} & \textbf{AdaBoost} & \textbf{KNN} \\ \hline
	\multirow{3}{*}{\textbf{\begin{tabular}[c]{@{}l@{}}Optimization in \\ test simulations (x)\end{tabular}}} & ALU                           & \multicolumn{8}{c}{4.55}                                                                                                        \\ \cline{2-10} 
	& ADC                           & \multicolumn{8}{c}{18.18}                                                                                                       \\ \cline{2-10} 
	& ECC                           & \multicolumn{8}{c}{7.14}                                                                                                        \\ \hline
	\multirow{3}{*}{\textbf{\begin{tabular}[c]{@{}l@{}}Optimization \\ in run-time (x)\end{tabular}}}         & ALU                           & 2.44            & 2.49           & 2.45           & 2.36         & 2.33        & 2.31        & 2.43              & 2.3          \\ \cline{2-10} 
	& ADC                           & 43.42           & 41.96          & 39.16          & 45.15        & 56.61       & 59          & 54.83             & 53.97        \\ \cline{2-10} 
	& ECC                           & 26.99           & 25.66          & 23.27          & 26.23        & 25.13       & 24.79       & 24.97             & 23.67        \\ \hline
	\multirow{3}{*}{\textbf{\begin{tabular}[c]{@{}l@{}}Coverage \\ regain (\%)\end{tabular}}}                 & ALU                           & 101.16          & 101.26         & 101.33         & 101.35       & 101.4       & 101.4       & 101.38            & 101.38       \\ \cline{2-10} 
	& ADC                           & 99.5            & 94             & 96.8           & 96.3         & 95.5        & 98.4        & 98.8              & 98.8         \\ \cline{2-10} 
	& ECC                           & \multicolumn{8}{c}{100}                                                                                                         \\ \hline
\end{tabular}
	}
	
	\vspace{-0.4cm}
\end{table}

\section{Conclusion}\label{sec:conclusion}

In this work, we developed a novel design-agnostic methodology that leverages the power of supervised machine learning and PyUVM-based testbenches to optimize coverage-driven verification. \ac{PyUVM} offered advantageous on-the-fly data collection for the \ac{ML} process. Our approach achieved significant reductions in test simulations, simulation cycles, and run-time. The automatic update of testbenches with new tests, sequences, and constraints enabled seamless integration of the \ac{ML} environment into the simulation environment, streamlining the verification process. Our results showed that linear models outperform other supervised regressors while achieving \SI{99}{\percent} coverage regain.

\printbibliography

\end{document}